\documentstyle[12pt]{article}
\begin{document}
\begin{flushright}
NTUA 45/94\\
May 94 \\
\end{flushright}
\begin{center}
{\bf TACHYON EFFECTS ON THE 2-DIM}\\
{\bf    BLACK HOLE GEOMETRY }\\
\vspace*{1cm}
{\bf G.A. Diamandis,B.C. Georgalas}\\
{\it Athens University,Physics Department,} \\
{\it Nuclear and Particle Physics Section,} \\
{\it Panepistimiopolis,Kouponia,} \\
{\it GR-157 71,Athens,Greece.}\\
\vspace*{0.3cm}
{\bf and}\\
\vspace*{0.3cm}
{\bf E. Papantonopoulos}\\
{\it National Technical University}\\
{\it Physics Department}\\
{\it GR-157 80 Zografou,}\\
{\it Athens,Greece.}\\
\vspace*{3cm}
{\bf ABSTRACT} \\
\end{center}
\noindent
We study solutions of the tree level string effective action in the
presence of the tachyon mode.We find that the 2-dim. static black hole
is stable against tachyonic perturbations.For a particular ansatz for
the tachyon field we find an exact solution of the equations of motion
which exhibits a naked singularity.In the case of static fields we
find numerically that the full system has a black hole solution,with
the tachyon regular at the horizon.
\vspace*{2cm}
\noindent
\thispagestyle{empty}
\vfill\eject
\setcounter{page}{1}

Black holes in two-dimensional gravity have received recently a lot of
attention, following Witten's identification of a black hole as a
string theory background \cite{Witten,Wadia,Verlinde}. The two-dimensional
dilaton gravity model was subsequently analyzed by many authors as a simple
laboratory to study black hole physics \cite{Giddings}.
The tachyon field, is a physical
string mode that appears in the effective two-dimensional string
theory, and the question of how the black hole geometry is affected by
the presence of the tachyon field is of interest by itself.  This
question was studied by a number of authors
\cite{Alwis,Rama,Peet,Minic} using various
approximation schemes, but without conclusive results as to whether
the tachyon back reaction destabilizes the black hole geometry.

De Alwis and Lykken \cite{Alwis}
found that a static tachyon field does not alter
the black hole character of the dilaton gravity solution.  They have
included in their investigation tachyon back reaction on the metric,
but their results are valid in the asymptotic regime where they can
ignore O($T^2$) terms.  Using a similar approximation, Rama
\cite{Rama}
found tachyon solutions but with a singular scalar curvature at the
horizon.  Non-singular static solutions were found by the authors of
ref. \cite{Peet}, by solving the tachyon equation in the dilaton gravity
background.

In this paper we undertake a more carefull and detailed study of the
tachyon effects to the dilaton gravity black hole solution.  We shall
first see that the black hole solution is stable against small
perturbations of the tachyonic field.
We will show that, if one considers the tachyon equation in a static
black hole background, then a small variation of the tachyon field by
$\delta T$ does not develop any growing mode, and therefore the black hole is
stable.

Of course a more interesting situation arises when tachyon back
reaction is considered.  In this case the main difficulty is the non
linear nature of the coupled differential equations of motion.
Working in light cone coordinates and assuming that the tachyon field
is a function of only $x^+$ or $x^-$ then an exact solution can be found.  This
is possible because the dilaton field decouples from the tachyon
equation which in return can be solved.  We attribute this decoupling
to a preservation of an $\rho-\phi$ symmetry  \cite{Russo}.
The characteristic of this solution is that the tachyon field is
singular at some finite $x^+$ and since the scalar curvature is
proportional to the tachyon field, it is also a physical singularity.
This singularity is a naked singularity, because as one can easily
see, there is no horizon to cover it.  It is interesting to note, that
this naked singularity remains stable against small perturbations of
the tachyonic field.

It is difficult to get any other exact solution for different ans\"atze
of the tachyonic field.  A considerable simplification arises if all
the fields are static.  In this case we end up with a simplest system
of coupled differential equations.
    This system can be studied numerically.  As we shall discuss in the
following, the numerical solution exhibits a clear black hole
behaviour with a tachyon field regular at the horizon.  The main
characteristic of this static solution is that the physical
singularity occurs at the strong coupling regime, where both the
tachyon and dilaton fields are singular.  Such a picture is supported
by an analytic expansion procedure in which the expansion parameter
plays the role of the mass in Callan et.al. \cite{Callan}.
In the case of $T=0$ in the first order in $\mu$ we get similar results
as in \cite{Alwis}, while from the second order in $\mu$ we start recognizing
the features of the numerical solution.

The action of the dilaton tachyon system coupled to gravity in two
dimensions is
 \begin{equation}
 S=\frac{1}{2\pi}
 \int d^2x\sqrt{-g}
 \{e^{-2\phi}
 [R+4(\nabla\phi)^2-(\nabla T)^2
 -V(T)+4\lambda^2 ] \}
 \label{eq:action}
 \end{equation}
 where V(T) is the tachyon potential.  There are ambiguities of the
tachyon potential in string theory \cite{Banks},
but in our discussion quadratic configurations for the tachyon fields
will be considered.   The equations of
motion resulting from the action (\ref{eq:action}) are
 \begin{equation}
 \Box\phi -(\nabla\phi)^2+\frac{1}{4}R
 -\frac{1}{4}(\nabla T)^2
 +\frac{1}{2} T^2+\lambda^2 =0
 \label{eq:dila}
 \end{equation}
 \begin{equation}
 \Box T-2(\nabla\phi)(\nabla T)+2T=0
 \label{eq:tach}
 \end{equation}
 \begin{equation}
 \Box\phi-2(\nabla\phi)^2
 +T^2+2\lambda^2=0
 \label{eq:metr}
 \end{equation}
 Using the dilaton eq.(\ref{eq:dila}) the metric equation
(\ref{eq:metr})
  becomes
 \begin{equation}
 \Box\phi+\frac{1}{2}R-\frac{1}{2}(\partial T)^2=0
 \label{eq:para}
 \end{equation}
  The above equations can be most easily analyzed if we work in the
conformal gauge
      \[g_{\mu\nu}=e^{2\rho}n_{\mu\nu}
      \quad ,
      \quad
      n_{\mu\nu}=diag(-1,1) \]
      Using also light-cone coordinates $x^\pm = x^0\pm x^1$, equations
(\ref{eq:dila}),(\ref{eq:tach}) and (\ref{eq:para}) become
  \begin{equation}
   \partial_+\partial_-\phi
   -(\partial_+\phi)(\partial_-\phi)
    -\frac{1}{2}\partial_+\partial_-\rho
    -\frac{1}{4}(\partial_+T)(\partial_-T)
    -\frac{1}{8}e^{2\rho}(T^2+2\lambda^2)=0
    \label{eq:dil1}
    \end{equation}
\begin{equation}
\partial_+\partial_-T-(\partial_+ \phi)(\partial_- T)
-(\partial_- \phi)(\partial_+ T)
-\frac{1}{2}e^{2\rho} T=0
\label{eq:tac1}
\end{equation}
       \begin{equation}
        \partial_+\partial_- \phi
        -\partial_+\partial_- \rho
        -\frac{1}{2}(\partial_+T)(\partial_-T)=0
        \label{eq:par1}
        \end{equation}
and in addition we have two constraints
\begin{equation}
\partial_\pm \partial_\pm \phi
-2 (\partial_\pm \phi)(\partial_\pm \rho)
-\frac{1}{2}(\partial_\pm T)(\partial_\pm T)=0
\label{eq:cons}
\end{equation}
In the case of $T=0$ we get the familiar Witten's solution which in the
conformal gauge is \cite{Callan}
\begin{equation}
e^{-2\phi}=e^{-2\rho}=\frac{M}{\lambda}
-\lambda^2 x^+ x^-
\label{eq:hole}
\end{equation}
and for the metric
\begin{equation}
   ds^2=-\frac{1}{2}(\frac{dx^+dx^-}
       {\frac{M}{\lambda}- \lambda^2 x^+ x^-})
\label{eq:line}
\end{equation}
where M is the black hole mass.  The singularity is at $x^+ x^- =
\frac {M} {\lambda ^3}$ and the horizon at $x^+ x^- = 0$.

In order to study the black hole stability under small tachyon
perturbations we define $\tilde{T}=e^{-\phi} T$ to make the tachyon kinetic
terms canonical, and assume
            \[\tilde{T}=\tilde{T}_0+\delta\tilde{T}\]
Then the tachyon equation (\ref{eq:tac1}) becomes
     \begin{equation}
      \partial_+ \partial_- (\delta\tilde{T})
       -[(\partial_+ \phi)(\partial_- \phi)
         -\partial_+\partial_- \phi
         +\frac{1}{2}e^{2^\rho}]
         (\delta \tilde{T}) =0
      \label{eq:wave}
      \end{equation}
Using the static solution (\ref{eq:hole}) and defining the space-time
coordinates
   \begin{eqnarray}
    r&=&ln(\frac{M}{\lambda} -\lambda^2 x^+ x^-) \nonumber\\
    t&=&-ln(-\frac{x^-}{x^+})
    \label{eq:rot}
    \end{eqnarray}
we get
  \begin{equation}
   \partial_t ^2 (\delta\tilde{T})
   -A^2{(r)} \partial_r^2 (\delta\tilde{T})
   -\frac{M}{\lambda} e^{-r} A{(r)} \partial_r (\delta \tilde{T})
    +\frac{1}{4\lambda} A{(r)}
   (\frac{\lambda^2 -2}{\lambda}
     +M e^{-r})
    (\delta \tilde{T}) =0
    \label{eq:schr}
    \end{equation}
with
     $ A{(r)}=(1-\frac{M}{\lambda}e^{-r} ) $.
Separating variables with
    $  \delta \tilde{T} =
       \tilde{T}(t)
       \tilde{R}(r) $,
then the space dependent equation becomes
    \begin{equation}
     \frac{d^2 \tilde{R}}{d \xi ^2}
      +(E^2-U(\xi)) \tilde{R} =0
    \label{eq:dia}
    \end{equation}
with $\xi=r+lnA$, and  $U(\xi)$ a complicated function of $\xi$.
For $\lambda^2\geq 2$ the potential is positive and this implies that
the corresponding eigenvalues E are real and positive, therefore, there are not
growing modes for the tachyon disturbances.  This means that the
static black hole is stable against small tachyon perturbations.In
\cite{Peet} an equation similar to eq.(\ref{eq:schr}) was solved
and a nonsingular solution
was found.  According then to our previous discussion this tachyon hair
is stable against small perturbations of the tachyon field.
Of course the stability found above is in agreement with the results
of \cite{Diamandis},where time dependent tachyon perturbations in the black
hole solution were discussed.
Nevertheless the value of the tachyon field is large near the
singularity  \cite{Wadia} ,
therefore the back reaction of the tachyon into the gravity dilaton
system cannot be ignored.

In an attempt to solve the equations (\ref{eq:dil1}) - (\ref{eq:cons})
 in order to include the back
reaction, we assume that the tachyon is a function of $x^+$.
Then we get the solution
     \begin{equation}
     e^{-2 \phi} = e^{-2 \rho}=u_+ +u_- -
     \frac{x^-}{2} \int ^{x^+} (T^2 +2 \lambda^2)
     \label{eq:sol}
     \end{equation}
where $u_-$ is linear in $x^-$
     \begin{equation}
     u_- = b_- + c_- x^-
     \label{eq:lin}
     \end{equation}
while the $u_+$ and the tachyon field obey the equations
         \begin{equation}
         \partial_+ ^2 u_+ +(u_+ + b_-)(\partial_+T)^2=0
         \label{eq:you}
         \end{equation}
    \begin{equation}
    T+\frac{1}{2} (\partial_+T)
   \int^{x^+} (T^2+2\lambda^2)-c_- (\partial_+T)=0
   \label{eq:man}
   \end{equation}
It is worthing to observe, that our choise of $T=T(x^+)$ (or similarly
$T=T(x^-)$) respects the $\rho-\phi$ symmetry \cite{Russo}
as we can immediatly see from eq.(\ref{eq:par1}).
A consequence of that,is the
"simple" form of the tachyon equation, where there are no metric or
dilaton dependent terms.  Eq.(\ref{eq:man}) can be solved and gives
       \begin{equation}
       \frac{d T^2}{d x^+}=
       -a (T^2)^{\frac{\lambda^2}{2} +1} e^{\frac{T^2}{4}}
       \label{eq:two1}
       \end{equation}
where $a$ is a constant which must be choosen positive for T to
vanish asymptotically.Eq.
(\ref{eq:you}) can be solved for $u_+$ using
the solution for $T^2$ from eq.(\ref{eq:two1}).
Substituting $u_+$ and $u_-$ to
eq.(\ref{eq:sol}) we finally get
     \begin{equation}
     e^{-2\phi}=e^{-2\rho}=
     C_1 F(1, \frac{\lambda^2}{2}+1 ; -\frac{T^2}{4}) +
     C_2 (-\frac{T^2}{4})^{-\frac{\lambda^2}{2}} e^{- \frac{T^2}{4}}
     -\frac{2 e^{-\frac{T^2}{4}}}
      {a (T^2)^{\frac{\lambda^2}{2}}} x^-
      \label{eq:geo}
      \end{equation}
where $C_1$ and $C_2$ arbitrary constants.  To simplify our discussion
we choose $\lambda ^2 = 2$,but our results are valid for every
$\lambda^2 \geq 2$. The tachyon eq.(\ref{eq:two1}) can be solved for
$x^+$ giving
      \begin{equation}
      x^+ = \frac{1}{4a} E_i (-\frac{T^2}{4}) +
      \frac{e^{-\frac{T^2}{4} }}
      {a T^2 } +d
      \label{eq:xip}
      \end{equation}
where d is a constant.  Using eq.(\ref{eq:xip}) we get for $e^{-2\phi}$
        \begin{equation}
        e^{-2\phi}=e^{-2\rho}=
        \frac{4C_1}{T^2}
        (1-e^{-\frac{T^2}{4}}) -2x^-
        \frac{1}{a T^2}
        e^{-\frac{T^2}{4}}
        \label{eq:tat}
        \end{equation}
where $x^-$ was shifted by a constant.  We can analyze now our results.
First of all from eq.(\ref{eq:xip}) we can see that the tachyon becomes
infinite
at some finite $x_0 ^+$.
This infinity is a
physical singularity as we can see calculating the curvature
\begin{equation}
R\propto \frac{C_1 T^2 }
{4C_1 (1-e^{-\frac{T^2}{4}}) - \frac{2}{a}x^-
e^{-\frac{T^2}{4}}}
\label{eq:curv}
\end{equation}
Of course there is also the usual physical singularity of $x^-=2a
C_1 (e^{\frac{T^2}{2}}-1) $ where the denominator of (\ref{eq:curv})
vanishes.The possible event horizon is given by the equations
$\partial _- \rho =\partial _+\rho=0$.  From
$\partial _+ \rho=0$ we get
$x^-=2a C_1 ( \frac {e^{\frac{T^2}{4}}} {1+\frac{T^2}{4}} -1)$
which covers the singularity at $x^-$,while $\partial_-\rho=0$ is satisfied
only when $T^2\rightarrow\infty$.  Therefore,the strong coupling
regime created by the tachyon infinity is a naked singularity
(see Fig.1).

The question now is, if this new solution we found is stable under
small perturbations of the tachyon field.  We follow a similar
procedure as in the static case. In the tachyon eq.(\ref{eq:wave})
 we substitute
the solution for $\phi$ from eq.(\ref{eq:tat}).In some appropriate variables x
and t we find the wave equation
       \begin{equation}
       (\partial_t ^2 -\partial_x ^2) (\delta \tilde{T})
       +V(x,t)(\delta \tilde{T})=0
       \label{eq:weq}
       \end{equation}
where V(x,t) is a complicated potential, a function of both x and t.
One can show that the potential remains positive and bounded for every
t, and therefore for $\delta \tilde{T}$ disturbance only oscillatory
modes are expected.

What we found is an exact solution which represents a naked
singularity.  Stability arguments tell us that this solution is stable
against small tachyon perturbations.The characteristic of this solution is
that it preserves an $\rho - \phi$ symmetry.  This symmetry is usefull
at the semi classical level because it helps in the solution of the
semi-classical dilaton equation of motion \cite{Bilal,Russo}.It would be
interesting to see of what happens to the geometry if we include to
our exact solution the one loop correction of the tachyon field
\cite{Future}.
Being unable to get any other exact solution,we restrict the dilaton field
to be static,that is a function of $x^+ x^-$.If we make the following
change of variables in the equations of motion
            \[x=ln(-x^+ x^- )\]
            \begin{equation}
            t=ln(-\frac{x^+}{x^-})
            \label{eq:var}
            \end{equation}
we find that the consistency of the system of eqs (\ref{eq:dil1})
- (\ref{eq:cons}) inforces all the other fields to be static and the
equations of motion become
        \begin{equation}
        T'' -2\phi' T' + \frac{1}{2}e^{2\rho+x} T=0
        \label{eq:stac}
        \end{equation}
\begin{equation}
\phi'' -2(\phi')^2 +\frac{1}{4}
e^{2\rho+x} (T^2+2\lambda^2)=0
\label{eq:sdil}
\end{equation}
         \begin{equation}
         \phi'' -\rho'' -\frac{1}{2} (T')^2 =0
         \label{eq:smet}
         \end{equation}
\begin{equation}
\phi'' -\phi' -2 \phi'\rho' -\frac{1}{2} (T')^2 =0
\label{eq:scon}
\end{equation}
Thus, in the tachyon equation there are explicit dilaton dependent
terms which make the solution very difficult.  On the other hand
equation (\ref{eq:smet}) tell us that the $\rho - \phi$ symmetry is not
maintained any more.Nevertheless we can solve the system
of eqs.(\ref{eq:stac}) -(\ref{eq:scon}) numerically,specifying the initial
conditions of the fields and their derivatives.To quarantee the
asymptotic flatness of the fields it is sufficient to start with a
negative value of the dilaton derivative.For a specific choise of the
initial conditions the results are given in Figs (2) and (3).

In Fig.(2) the dilaton field is plotted as a function of the
variable x.  We see that from the value $\phi=-\infty$,in which it
tends linearly in the asymptotic region,is monotonically increasing
becoming singular at finite x.The behaviour of the tachyon field is
shown in Fig.(3).Asymptotically it tends to zero and it becomes
singular at the same x,where the dilaton is infinite.This point is the
position of the physical singularity.Comparing eqs (\ref{eq:smet}) end
(\ref{eq:scon}) we get
    \begin{equation}
    \rho' + \frac{1}{2}=\mu e^{2\phi}
    \label{eq:hor}
    \end{equation}
where $\mu$ is a positive intergration constant.Thus $\rho'$ varies
monotonically from $+\infty$ to $-\frac{1}{2}$.This means that it
vanishes only once at some finite x where both the dilaton and the
tachyon fields are regular.Since for the static case $\rho'=0$ solves
the equations $\partial_+\rho=\partial_-\rho=0$, this point gives the
position of the horizon.

Such a picture is supported by a more formal analysis of the eqs
(\ref{eq:stac}) - (\ref{eq:scon}).
Assume that the dilaton field is a monotonic function of x. Then we can
define a function $p(\phi)$ as the first derivative of $\phi$,
$\frac{d\phi}{dx}=p(\phi)$.Defining
 $\xi=\rho'+\frac{1}{2}$
 and using eq.(\ref{eq:hor}) the system of equations
 (\ref{eq:stac})-(\ref{eq:scon}) become
     \begin{equation}
     p^2 \frac{d^2T}{d\phi^2} +
     p \frac{dp}{d\phi} \frac{dT}{d\phi}
     -2p^2 \frac{dT}{d\phi}
     +\frac{1}{2} e^{2\xi} T=0
     \label{eq:expa}
     \end{equation}
\begin{equation}
4(2p^2  -p\frac{dp}{d\phi} )
-(T^2+2\lambda^2) e^{2\xi}=0
\label{eq:expb}
\end{equation}
             \begin{equation}
             p\frac{d\xi}{d\phi}-\mu e^{2\phi}=0
             \label{eq:expc}
             \end{equation}
\begin{equation}
\frac{dp}{d\phi} -2 \mu e^{2\phi}
-\frac{1}{2} p (\frac{dT}{d\phi})^2=0
\label{eq:expd}
\end{equation}
We are looking for solution analytic in $\mu$
     { \[p=p_0 +\mu p_1 +\mu^2p_2+...\] }\\
     {  \[\xi=\xi_0 +\mu \xi_1 +\mu^2 \xi_2 +...\] }\\
     {  \[T=T_0 +\mu T_1 +\mu^2 T_2 +...\] }\\
This can be justified by the fact that the Witten's solution is
analytic in the black hole mass which is related to the parameter
$\mu$ in eq.(\ref{eq:hor}).It is easy to see that the zero order solution
is the flat space solution with $T_0 =0$.The first order system is
        \begin{equation}
        \xi_1 '=-2e^{2\phi}
        \quad ,
        \quad
        p_1 '=2e^{2\phi}
        \label{eq:firs}
        \end{equation}
\begin{equation}
T_1 '' -2T_1 '+\frac{2}{\lambda^2} T_1=0
\label{eq:first}
\end{equation}
Eq.(\ref{eq:first}) is the linearized tachyon equation which was considered in
\cite{Alwis}.  Its solution for $\lambda ^2=2$ is
         \begin{equation}
         T_1=(A_1 +B_1 \phi) e^\phi
         \label{eq:rela}
         \end{equation}
where  $A_1$ and $B_1$ are arbitraty constants.
The second order solution gives a tachyon of the form
          \begin{equation}
          T_2=(A_2+ B_2 \phi)e^\phi
          +(\Gamma_2 +\Delta_2 \phi)e^{3\phi}
          \label{eq:fani}
          \end{equation}
with $A_2$, $B_2$, $\Gamma_2$, $\Delta_2$ constants.
Therefore the tachyon goes to infinity when $\phi$ gets infinite.
In can be shown that at any order the tachyon is always a combination
of exponentials $e^{\alpha_i \phi}$ with $\alpha_i >0$.
Furthermore the second order solution for the dilaton gives a physical
singularity at a finite point,yielding the basic characterists of the
numerical solution we have found.

In conclusion, our main result is that in the case of static fields,
the tachyon back reaction maintains the black hole structure of the
gravity dilaton system.  We have seen that by a numerical analysis of
the full system without any approximation, while this is also
supported by an expansion procedure where we have assume that the
the dilaton field is a monotonic function. In our solution,
the tachyon is regular at the horizon, while the physical singularity
occurs where both the dilaton and tachyon fields become infinite.Also
for a non-static tachyon, in a limited case of a particular
anzatz, we found an exact solution which represents a naked
singularity.Finally, when the tachyon field is not back reacted
we found that the black hole geometry is stable against tachyon
perturbations.
\vspace{1cm}

We had benefited by the discussions we had with A.Lahanas in the
early stages of this work.We also want to thank C.Bachas,E.
Kiritzis,N.Mavromatos and L. Thorlacius for valuable discussions.
Work partially supported dy C.E.E. Science Program SCI-CT92-0792

\end{document}